\documentclass[11pt]{article}
\usepackage{amsmath,amssymb, mcite}
\usepackage{stmaryrd}
\usepackage{rotating}
\usepackage{caption}

\newcommand{\tM}{{\tt M}}
\newcommand{\tN}{{\tt N}}
\newcommand{\tP}{{\tt P}}
\newcommand{\tQ}{{\tt Q}}
\newcommand{\tR}{{\tt R}}
\newcommand{\tS}{{\tt S}}
\newcommand{\tT}{{\tt T}}
\newcommand{\tU}{{\tt U}}

\newcommand{\cV}{{\cal V}}
\newcommand{\cM}{{\cal M}}
\newcommand{\cN}{{\cal N}}

\newcommand{\cB}{{\cal B}}
\newcommand{\cD}{{\cal D}}
\newcommand{\cQ}{{\cal Q}}
\newcommand{\cP}{{\cal P}}
\newcommand{\cR}{{\cal R}}
\newcommand{\cS}{{\cal S}}
\newcommand{\cT}{{\cal T}}

\newcommand{\cL}{{\cal L}}

\newcommand{\bR}{{\bf{R}}}
\newcommand{\bGa}{{\bf{\Gamma}}}
\newcommand{\bT}{{\bf{T}}}

\begin{document}

\begin{titlepage}
\hfill DAMTP-2014-9\hspace*{1.7mm}

\hfill  AEI-2014-000\hspace*{4.2mm}
\vspace{2.5cm}
\begin{center}

{{\LARGE  \bf Einstein-Cartan Calculus for Exceptional Geometry }} \\

\vskip 1.5cm {Hadi Godazgar$^{\star}$ 
, Mahdi Godazgar$^{\dagger}$ and
Hermann Nicolai$^{\ddagger}$}
\\
{\vskip 0.5cm
$^{\star \dagger}$DAMTP, Centre for Mathematical Sciences,\\
University of Cambridge,\\
Wilberforce Road, Cambridge, \\ CB3 0WA, UK\\
\vskip 0.5cm
$^{\ddagger}$Max-Planck-Institut f\"{u}r Gravitationsphysik, \\
Albert-Einstein-Institut,\\
Am M\"{u}hlenberg 1, D-14476 Potsdam, Germany}
{\vskip 0.35cm
$^{\star}$H.M.Godazgar@damtp.cam.ac.uk, $^{\dagger}$M.M.Godazgar@damtp.cam.ac.uk, 
$^{\ddagger}$Hermann.Nicolai@aei.mpg.de}
\end{center}

\vskip 0.35cm

\begin{center}
\today
\end{center}

\noindent

\vskip 1.2cm

\begin{abstract}
\noindent 
In this paper we establish and clarify the link between the recently found E$_{7(7)}$ generalised 
geometric structures, which are based on the SU(8) invariant reformulation 
of $D=11$ supergravity proposed long ago,  and newer results obtained in the 
framework of recent approaches to generalised geometry, where E$_{7(7)}$ duality is 
built in and manifest from the outset. 
In making this connection, the so-called generalised vielbein postulate plays a key
role. We explicitly show how this postulate can be used  to define an E$_{7(7)}$ valued 
affine connection and an associated covariant derivative, which yields a generalised 
curvature tensor for the E$_{7(7)}$ based exceptional geometry.  The analysis of
the generalised vielbein postulate also provides a natural explanation for the 
emergence of the embedding tensor from higher dimensions.
\end{abstract}

\end{titlepage}

\section{Introduction}

Recent progress \cite{GGN13}, along the lines of an older proposal~\cite{dWNsu8}, 
on understanding the extent to which the E$_{7(7)}$ Cremmer-Julia duality symmetry \cite{cremmerjulia, so(8)} is inherent to the \emph{full} $D=11$ supergravity theory \cite{CJS} has lead to a new formulation of the $D=11$ theory, which apart from pointing to new geometric structures in eleven dimensions, provides an appropriate framework in which to address questions regarding the relation between $D=11$ supergravity and four-dimensional maximal gauged supergravity theories \cite{KKdual, GGNss}. In this paper, we will
clarify the relation of these results to more recent approaches to generalised geometry,
especially \cite{CSW, BCKT, HO}, and show how a synthesis  of the different approaches emerges.

The formalism of Ref.~\cite{GGN13} is based on the SU(8) invariant reformulation of $D=11$ supergravity \cite{dWNsu8}, in which the local and global gravitational symmetries of the eleven-dimensional theory are abandoned and one performs a $4+7$ split of all fields in the theory.  Importantly, dependence on all eleven coordinates is retained throughout and one remains 
on-shell equivalent to the original theory throughout the construction. An essential 
characteristic of the analysis of Ref.~\cite{dWNsu8}, and a main distinguishing feature 
in comparison with more recent work, is the use of supersymmetry transformations to 
find new SU(8) and E$_{7(7)}$ structures in the eleven-dimensional theory.  The most 
significant such structures are the ``generalised vielbeine'' \cite{dWNsu8, dWN13, GGN13}, which replace the eleven-dimensional fields that would contribute to scalar degrees of freedom in a reduction to four dimensions. As in \cite{dWNsu8}, these are derived by considering the supersymmetry transformation of eleven-dimensional fields that would contribute to vector degrees of freedom in a reduction to four dimensions.  A crucial ingredient in constructing 
the full set of ``generalised vielbeine'' is to consider {\em dual fields in eleven dimensions}.  
These building blocks are to be viewed as the components of a single E$_{7(7)}$ 56-bein $\cV$ that we shall henceforth simply refer to as {\em the} ``generalised vielbein",  in analogy with the terminology used in more recent literature \cite{BGPW}.  In particular, the generalised vielbein as derived directly from the $D=11$ theory in \cite{GGN13} coincides with the generalised vielbein that lies at the heart of other recent approaches to generalised geometry \cite{hillmann} (see also \cite{BGPW}), where it is constructed from the E$_{7(7)}$/SU(8) coset using an algebraic 
method known as non-linear 
realisation \cite{BO, west2000, locale11}. More recently, the generalised geometry 
ideas that have been used to describe the seven-dimensional sector of $D=11$ supergravity 
in a $4+7$ split have been extended to incorporate the four-dimensional part, in this
way arriving at an E$_{7(7)}$ covariant extension of the whole theory \cite{Hohm:2013pua,HO}. 

An important aspect of the formalism developed in \cite{GGN13} is the fact that the components of the generalised vielbein satisfy differential constraints \cite{dWNsu8, GGN13} -- called ''generalised vielbein postulates" (GVPs) due to their resemblance to the usual vielbein postulate in differential geometry.  It should be emphasised that here these equations are not postulated,
but follow directly from the explicit expressions for the generalised vielbein in terms
of the various fields and dual fields of $D=11$ supergravity. In this sense, the present 
approach is `bottom up', in contrast to other approaches, where similar relations
follow from more abstract geometrical reasoning. One of our main results here is to
show how these ingredients can be used to develop an Einstein-Cartan calculus
that is largely analogous to the one for the standard vielbein.

The GVPs divide into two sets: those in which the derivative acting on the component of the generalised vielbein is taken with respect to the $D=4$ directions and those in which the derivative is with respect to the $D=7$ directions.  Using a terminology where ``external'' refers to $D=4$ in the $4+7$ split of $D=11$, and ``internal'' refers to $D=7$, even though we remain on-shell equivalent to the $D=11$ theory and no reduction is assumed, we refer to the former set as ``external GVPs" and the latter set as ``internal GVPs".
The GVPs are important in establishing a link between the $D=11$ theory and $D=4$ maximal gauged theories derived as a reduction thereof.  In particular, the external GVPs can be regarded as providing a higher dimensional origin of the embedding tensor \cite{NSmaximal3, NScomgauge3, dWSTlag, dWSTgauge, dWSTmax4}, as has been explicitly demonstrated for the $S^7$ reduction \cite{KKdual} and Scherk-Schwarz flux compactifications \cite{GGNss}. The relationship between $D=11$ supergravity and $D=4$ supergravity is an important aspect of the SU(8) invariant reformulation of the $D=11$ theory \cite{dWNsu8}, and recent developments therefrom \cite{dWN13, GGN13}, in, for example, establishing non-linear ans\"atze \cite{dWNW, dWN13, GGN, KKdual} and consistency of the $S^7$ reduction \cite{dWNconsis, NP}. Very recently, this aspect has also been studied in Ref.~\cite{LSW} where the generalised vielbein is related by a generalised Scherk-Schwarz ansatz to the E$_{d(d)}$ matrix parametrised by the scalars of maximal 
gauged supergravity. This allows them to verify/conjecture non-linear ans\"atze for various sphere reductions. The validity of the new ans\"atze can be established by an analysis along the lines of Refs.~\cite{dWNsu8, dWNW, dWN13, KKdual} for the appropriate sphere reductions.       

In this paper, we return to the reformulation of $D=11$ supergravity developed in \cite{GGN13} and proceed to make concrete the indications that there is an E$_{7(7)}$ generalised geometry underlying the constructions there.  In particular, we make contact with recent results in duality-manifest based approaches to generalised geometry \cite{CSW, BCKT}~\footnote{For further references see \cite{hullgenm, PW, BP, CSW2, Aldazabal:2013mya, Cederwall:2013naa, charlie, Aldazabal:2013via}.} that have focused on similar issues from a duality group perspective.  We condense all the objects and equations, in particular the GVPs, into an E$_{7(7)}$ covariant form such that the previous expressions can be obtained as particular components of the new expressions under SL(8) and GL(7) decompositions of E$_{7(7)}$.  Thus, even though general covariance in $D=11$ has been abandoned in the $4+7$ split, we obtain a reformulation that has general covariance in the $D=4$ directions and a ``generalised general covariance'' based on E$_{7(7)}$ in the $D=7$ space in a manner consistent with the results of Ref.~\cite{CSW, BCKT, HO}. 

A prerequisite for introducing E$_{7(7)}$ covariance, and thus replacing GL(7) indices 
with E$_{7(7)}$ indices, is that the seven-dimensional space on which the generalised 
geometry is constructed apparently requires an extension to 
a 56-dimensional space~\footnote{More precisely, the 11-dimensional space-time 
 manifold would have to be extended to a (4+56)-dimensional space, but we can 
 ignore the dependence on the 
four external coordinates for the argument to be presented.} such that the 
seven internal coordinates $\{y^m\}$ are extended to a set of 56 internal coordinates
$\{ y^\cM\}$, where $\cM$ labels the $\bf{56}$ representation of E$_{7(7)}$ \cite{hillmann}.
However, in order for the geometric structures, such as the algebra of generalised diffeomorphisms, to be consistent one must impose a constraint, the section condition, that ultimately reduces the enlarged space to an at most seven-dimensional 
space \cite{CSW, BCKT}. 
While the necessity of such a restriction is plainly evident from the fact that no consistent supergravity appears to exist beyond eleven dimensions, its necessity can also be seen
from a more geometrical perspective: supposing that the generalised vielbein $\cV$ {\em did} 
depend on 56 internal coordinates, we would have the textbook formula
$$
\cV_\cM (y)  \, = \, \cV'_\cN(y')\,  \frac{\partial y'^\cN}{\partial y^\cM}
$$
for the transformation under arbitrary diffeomorphisms in 56 dimensions. However, the 
transition matrix $\partial y'^\cM/ \partial y^\cN$ being an element of GL(56),  this operation
would throw the 56-bein $\cV$ out of the coset E$_{7(7)}/$SU(8). One might therefore ask
whether there exists a set of {\em restricted} diffeomorphisms in 56 dimensions, such that
$\partial y'^\cM/ \partial y^\cN  \in$ E$_{7(7)}$ and the transformed 
generalised vielbein remains in the coset. However, this possibility is excluded by 
Cartan's  Theorem, according to which there do not exist  `exceptional algebras
of vector fields' on manifolds, the only possibilities being (essentially) the algebras of ordinary
diffeomorphisms, volume preserving diffeomorphisms and 
symplectomorphisms \cite{cartan, ss} (see also Ref.~\cite{kac}).
Similar comments apply to the 3+8 split associated to the E$_{8(8)}$ duality
group, as already noted in \cite{KNS}.

In section \ref{prelim}, we review the required results from \cite{GGN13}, rewriting them in a manner that makes their E$_{7(7)}$ structure manifest. We rewrite the GVPs, in section \ref{sec:GVP}, using the E$_{7(7)}$ structures defined in section \ref{prelim}. Then, in section \ref{sec:gendiff}, we explicitly demonstrate how the coordinate and gauge transformations of the generalised vielbein can be packaged into a single transformation given by generalised diffeomorphisms \cite{CSW}. Finally, in section \ref{sec:gvps}, we similarly package the GVPs into single E$_{7(7)}$ covariant equations. The equation corresponding to the external GVPs is precisely of the same form as the Cartan equation in four-dimensional maximal gauged theories, allowing us to identify the higher dimensional object, an operator, that gives the embedding tensor upon reduction to four dimensions. On the other hand, the internal GVP is the generalised geometric analogue of the vielbein postulate and yields the generalised connection 
for the generalised geometry. We give the transformation properties of the generalised connection. Furthermore, we find that a covariant derivative defined using the generalised connection transforms as a generalised tensor density of weight $1/2$ less than the weight of the generalised tensor on which the covariant derivative acts. Thus, a generalised Riemann curvature tensor obtained by commuting two covariant derivatives transforms as a generalised tensor density of weight $-1.$ We explicitly present the components of the generalised Riemann tensor and note that it is indeed generalised gauge covariant.    

The conventions used in this paper are the same as those of Ref.~\cite{dWNsu8}. In particular, $M, N, \ldots$ and $A, B, \ldots$ denote eleven-dimensional spacetime and tangent space indices, respectively.  Indices $A, B, \ldots$ are also used as SU(8) indices.  However, it should be clear from the context what type of index is being referred to.  Similarly, $\mu, \nu, \ldots$ and $\alpha, \beta, \ldots$, and $m, n, \ldots$ and $a, b, \ldots$ denote $D=4$ and $D=7$ spacetime and tangent space indices, respectively.

\section{Preliminaries} \label{prelim}

\subsection{Generalised vielbein}

As explained in much detail in our previous work \cite{GGN13}, a generalised vielbein 
$\cV$, which can be viewed as a 56-bein of E$_{7(7)}$, can be defined 
{\em directly} in eleven dimensions. This 56-bein depends 
on the fields {\em and} on the  dual fields of $D=11$ supergravity, as obtained by performing a $4+7$ split on the original fields,  with all fields still depending on all 
eleven-dimensional coordinates.  In particular, it depends on the siebenbein $e_m{}^a$,
which is obtained from a $4+7$ decomposition of the original elfbein of $D=11$ 
supergravity in a triangular gauge (which breaks the original tangent space symmetry
SO(1,10) of the theory to SO(1,3) $\times$ SO(7)):
\begin{equation}\label{EMA}
E_M{}^A (x,y) \, = \,
\left(\begin{matrix} \Delta^{-1/2} e'_\mu{}^\alpha  & B_\mu{}^m e_m{}^a \\[2mm]
0 & e_m{}^a \end{matrix}\right)\;\; ,\qquad \Delta \equiv \det e_m{}^a.
\end{equation}
Here, as usual, we split the eleven-dimensional coordinates $\{z^M\}$ into four
external coordinates $\{x^\mu\}$ and seven internal coordinates $\{y^m\}$.
The 3-form and 6-form gauge fields, on which the 56-bein also depends, 
are linked via the duality relation
\begin{align} \label{F7}
 F_{M_1 \cdots M_7} & = 7! D_{[M_1} A_{M_2 \cdots M_7]} + 7! \frac{\sqrt{2}}{2}  
 A_{[M_1 M_2 M_3} D_{M_4} A_{M_5 M_6 M_7]} \notag \\
 &\hspace*{40mm}- \frac{\sqrt{2}}{192} i \epsilon_{M_1 \cdots M_{11}} \left(\overline{\Psi}_{R} \tilde{\Gamma}^{M_8 \cdots M_{11} RS} \Psi_{S} + 12 \overline{\Psi}^{M_8} \tilde{\Gamma}^{M_9 M_{10}} \Psi^{M_{11}} \right)
\end{align}
in eleven dimensions, from which all pertinent relations linking the 4-form and 7-form 
field strengths can be obtained by choosing the indices appropriately. Although 
we will ignore the fermionic terms in this duality relation in the remainder, it should be clear that
this duality relation introduces a hidden dependence of the 56-bein (which we are
about to present) on the fermionic fields as well.

A main result of \cite{GGN13} is thus the complete identification of the 56-bein in terms 
of the siebenbein, and the internal components of the 3-form and the 6-form,
such that $\cV \equiv \cV(e, A^{(3)}, A^{(6)})$. With proper E$_{7(7)}$ normalisation, 
the components of the generalised vielbein are explicitly given by
\begin{align}
\cV^{m}{}_{AB} &=  - \frac{\sqrt{2}}8 \Delta^{-1/2} \Gamma^m_{AB}, \label{gv11} \\[3mm]
\cV_{mn}{}_{AB} &= - \frac{\sqrt{2}}{8}  \Delta^{-1/2} \left(
\Gamma_{mn}{}_{AB} + 6 \sqrt{2} A_{mnp} \Gamma^{p}_{AB} \right),
\label{gv12} \\[3mm]
\cV^{mn}{}_{AB} & = - \frac{\sqrt{2}}8 \cdot \frac1{5!} \, \eta^{mnp_1\cdots p_5}
   \Delta^{-1/2} \Bigg[
\Gamma_{p_1 \cdots p_5}{}_{AB} + 60 \sqrt{2} A_{p_1 p_2 p_3} \Gamma_{p_4
p_5}{}_{AB}  \notag \\
 & \hspace*{50mm}  - 6! \sqrt{2} \Big( A_{q p_1 \cdots p_5} - 
\frac{\sqrt{2}}{4} A_{qp_1 p_2} A_{p_3 p_4 p_5} \Big) \Gamma^{q}_{AB}
\Bigg], \label{gv13} \\[2mm]
\cV_{m\, AB} &=
- \frac{\sqrt{2}}8 \cdot\frac1{7!} \, \eta^{p_1\cdots p_7}
\Delta^{-1/2} \Bigg[
(\Gamma_{p_1 \cdots p_7} \Gamma_{m}{})_{AB} + 126 \sqrt{2}\ A_{m p_1
p_2} \Gamma_{p_3 \cdots p_7}{}_{AB} \notag \\
 & \hspace*{37mm}  + 3\sqrt{2} \times 7! \Big( A_{m p_1 \cdots p_5} +
\frac{\sqrt{2}}{4} A_{m p_1 p_2} A_{p_3 p_4 p_5} \Big) \Gamma_{p_6
p_7}{}_{AB} \notag \\[2mm]
  & \hspace*{42.5mm} + \frac{9!}{2} \Big(A_{m p_1 \cdots p_5} +
\frac{\sqrt{2}}{12} A_{m p_1 p_2} A_{p_3 p_4 p_5} \Big) A_{p_6 p_7 q }
\Gamma^{q}{}_{AB} \Bigg], \label{gv14}
\end{align}
where $\Gamma^m \equiv e^m{}_a \Gamma^a$ are the $D=7$ gamma matrices with seven-dimensional curved indices and $\eta^{m_1 \dots m_7}$ is the seven-dimensional permutation symbol (tensor density of weight +1). These expressions are obtained by insisting on the E$_{7(7)}$ covariance of the supersymmetry variations (after appropriate field redefinitions), 
see also remarks in section~2.2 below.

The vielbein is subject to local SU(8) rotations (depending on all eleven coordinates),
such that the above expressions in terms of quantities of $D=11$ supergravity
correspond to a special gauge choice, as explained already in \cite{dWNsu8}.
Furthermore, complex conjugation raises (lowers) SU(8) indices
\begin{equation}
\cV_{\tM\tN}{}^{AB} \equiv (\cV_{\tM\tN\, AB})^*  \;\; , \quad
\cV^{\tM\tN AB} \equiv (\cV^{\tM\tN}{}_{AB})^*,
\end{equation}
where we have combined the GL(7) indices $m,n,\dots$ into SL(8) indices 
$\tM,\tN,\dots$ according to~\footnote{For brevity, we will often  use the 
simplifying notation $\cV^{m} \equiv \cV^{m8} = - \cV^{8m}$ and 
$ \cB^{m} \equiv \cB^{m8} = - \cB^{8m}, etc.$}
\begin{equation}
\cV_{\tM\tN} \equiv \big( \cV_{mn} , \cV_{m8}  \big) \;\; , \quad
\cV^{\tM\tN} \equiv \big( \cV^{mn} , \cV^{m8}  \big).
\end{equation}
That is, complex conjugation only affects the SU(8) indices. We will also use 
proper E$_{7(7)}$ indices $\cM, \cN,\dots$ corresponding to the $\bf{56}$
representation, such that
\begin{equation}
\cV_\cM \equiv \big(\cV_{\tM\tN} , \cV^{\tM\tN}\big)  \;\; , \quad
\cV^\cM = \Omega^{\cM\cN} \cV_\cN \equiv
\big(\cV^{\tM\tN} , - \cV_{\tM\tN}\big)  \;,
\end{equation}
where the components of the symplectic form $\Omega^{\cM\cN}$ are
\begin{align}
 \Omega^{\tM \tN}{}_{\tP \tQ}&=\delta^{\tM \tN}_{\tP \tQ}, \qquad \Omega_{\tM \tN}{}^{\tP \tQ}=-\delta_{\tM \tN}^{\tP \tQ}, \notag \\
 \Omega_{\tM \tN\, \tP \tQ}&=0, \qquad \hspace{3mm} \Omega^{\tM \tN\, \tP \tQ}=0,
\end{align}
and $\Omega_{\cM \cN}$ is given by $\Omega^{\cM\cP} \Omega_{\cN \cP} = \delta^{\cM}_{\cN}$.
Moreover, with the above normalisation, $\cV$ satisfies the E$_{7(7)}$ properties
\begin{align}
 \cV_{\cM}{}^{AB} \cV_{\cN\, AB} - \cV_{\cM\, AB} \cV_{\cN}{}^{AB} &= i\, \Omega_{\cM \cN}, \notag \\[2mm]
 \Omega^{\cM \cN} \cV_{\cM}{}^{AB} \cV_{\cN\, CD} &= i\, \delta^{AB}_{CD}, \notag \\[2mm]
 \Omega^{\cM \cN} \cV_{\cM}{}^{AB} \cV_{\cN}{}^{CD} &= 0,
\end{align}
which can be directly verified from definitions \eqref{gv11}--\eqref{gv14}.  The generalised vielbein also satisfies the following E$_{7(7)}$ covariant supersymmetry transformation
\begin{equation}\label{deltaV}
 \delta \cV_{\cM \, AB} = \sqrt{2} \Sigma_{ABCD} \cV_{\cM}{}^{CD},
\end{equation}
with the complex self-dual SU(8) tensor
\begin{equation}
 \Sigma_{ABCD} = \bar{\varepsilon}_{[A} \chi_{BCD]} + \textstyle{\frac{1}{4!}} \epsilon_{ABCDEFGH} \bar{\varepsilon}^{E} \chi^{FGH}. 
\end{equation}
In the SU(8) invariant reformulation, the $D=11$ gravitino $\Psi_{M}$ is rewritten in terms of SU(8) covariant chiral fermions $\varphi_{\mu}{}^{A}$ and $\chi_{ABC}$ and their complex conjugates $\varphi_{\mu}{}_{A}, \chi^{ABC}$ \cite{so(8), dWNsu8}. The precise relation between
\eqref{deltaV} and the $D=11$ supersymmetry variations also involves an SU(8) rotation
that we have dropped.

In addition to local SU(8) transformations, the generalised vielbein is subject to 
several gauge transformations which it inherits from the fields on 
which it depends, to wit, internal diffeomorphisms, and the tensor gauge
transformations associated to the 3-form and 
the 6-form gauge potentials. Recall that in our scheme all transformation parameters
depend on eleven coordinates. The transformations under internal diffeomorphisms
are straightforward to obtain:
\begin{align}
\delta \cV^{m}{}_{AB} &= \xi^p \partial_p \cV^{m}{}_{AB} \,-\,  \partial_p \xi^m \cV^{p}{}_{AB}
                    \, -\, \frac12 \partial_p \xi^p \cV^{m}{}_{AB}, \notag  \\[2mm]
\delta \cV_{mn\, AB} &= \xi^p \partial_p \cV_{mn\, AB} \,-2\, \partial_{[m} \xi^p \cV_{n]p\, AB}
                    \, -\, \frac12 \partial_p \xi^p \cV_{mn\, AB}, \notag \\[2mm]                     
\delta \cV^{mn}{}_{AB} &= \xi^p \partial_p \cV^{mn}{}_{AB} \,+2\, \partial_p \xi^{[m} \cV^{n]p}{}_{AB}
                    \, + \, \frac12 \partial_p \xi^p \cV^{mn}{}_{AB}, \notag \\[2mm]                     
\delta \cV_{m\, AB} &= \xi^p \partial_p \cV_{m\, AB} \, +\,  \partial_{m} \xi^p \cV_{p\, AB}
                    \, + \, \frac12 \partial_p \xi^p \cV_{m\, AB}. \label{Vdiff}
\end{align}
Note that the density terms come from the overall factor of $\Delta^{\pm1/2}$ in the definition of $\cV_{\cM}$.
With respect to the tensor gauge transformations, we have
\begin{equation} \label{2formgauge}
\delta A_{mnp} =  3! \, \partial_{[m} \xi_{np]} \;\; , \quad
 \delta A_{mnpqrs} =  3\sqrt{2} \, \partial_{[m} \xi_{np} A_{qrs]}  
\end{equation}
and 
\begin{equation} \label{5formgauge}
\delta A_{mnp} = 0  \;\: , \quad
\delta A_{mnpqrs} =  6! \, \partial_{[m} \xi_{npqrs]} 
\end{equation}
with the 2-form and 5-form gauge parameters $\xi_{mn}$ and $\xi_{mnpqr}$, respectively.
Substituting these transformations into the explicit expressions for the generalised vielbein components in \eqref{gv11}--\eqref{gv14}, it is straightforward to deduce
the transformation properties 
\begin{align}
&\delta \cV^{m}{}_{AB}= 0, \hspace{58mm} \delta \cV_{mn\, AB} = 36 \sqrt{2} \,\partial_{[m} \xi_{np]} \, \cV^{p}{}_{AB}, \notag \\[2mm]
&\delta \cV^{mn}{}_{AB} = 3\sqrt{2}\, \eta^{mnpqrst} \partial_{p} \xi_{qr} \,\cV_{st\, AB}, \hspace{18mm}
\delta \cV_{m\, AB} = 18 \sqrt{2} \, \partial_{[m} \xi_{np]} \, \cV^{np}{}_{AB}, \label{dV2form}
\end{align}
and 
\begin{align}
&\delta \cV^{m}{}_{AB} = \delta \cV_{mn\, AB} = 0, \hspace{20mm}
\delta \cV^{mn}{}_{AB} = - 6\cdot 6! \sqrt{2} \, \eta^{mnp_1\cdots p_5} 
\partial_{[q} \xi_{p_1 \cdots p_5]} \cV^{q}{}_{AB}, \notag \\[2mm]
&\delta \cV_{m\, AB} = 3\cdot 6! \sqrt{2} \, \eta^{n_1\cdots n_7} \partial_{[m} \xi_{n_1\cdots n_5]} 
   \cV_{n_6 n_7}{}_{AB}.  \label{dV5form}
\end{align}
We already see here that these transformation parameters can be nicely combined as
\begin{equation}
\Lambda^\cM \,\equiv\, \big( \Lambda^m, \Lambda_{mn}, \Lambda^{mn} , 0 \big)
\end{equation}
where $\Lambda^m \sim \xi^m \,,\, \Lambda_{mn} \sim \xi_{mn}$ and
$\Lambda^{mn} \sim \eta^{mnp_1\cdots p_5} \, \xi_{p_1\cdots p_5}$ (the precise coefficients
will be conveniently chosen later). In this way ordinary diffeomorphisms and tensor 
gauge transformations are unified into a single set of transformations.  This will be 
shown explicitly in section \ref{sec:gendiff}, where we will consider generalised 
diffeomorphisms and show how the above transformations can be compactly written 
in terms of a single generalised Lie derivative, see equation \eqref{genLie}. The `missing' seven 
components $\Lambda_m$ in this identification are obviously associated with 
`dual' internal  diffeomorphisms, but will actually be seen to drop out.

\subsection{Vector fields}
The components of the generalised vielbein can be obtained by considering the supersymmetry of a set of eleven-dimensional fields with one $D=4$ index \cite{dWNsu8, dWN13, GGN13}.  As such they are known as vectors in accord with the convention of using four-dimensional language for analogous $D=11$ structures adopted here. We similarly combine the vectors
into a {\bf56} of E$_{7(7)}$
\begin{equation}
\cB_{\mu}^{\cM} = (\cB_{\mu}^{\tM \tN},\, \cB_{\mu \, \tM \tN}).
\end{equation}
The proper definitions of these 56 vector fields follow from the identifications
\begin{align}
  {\cB_{\mu}}^{m} &= -\frac{1}{2} {B_{\mu}}^m, \hspace{30.5mm}
\cB_{\mu\, mn} = -3\sqrt{2}\, \big(A_{\mu mn} - B_\mu{}^p A_{pmn}\big), \notag \\[4mm]
{\cB_{\mu}}^{m n} &= -3\sqrt{2}\, {\eta}^{mnp_1\dots p_5} 
\left( A_{\mu p_1 \cdots p_5} - B_\mu{}^q A_{qp_1\cdots p_5}  -
       \frac{\sqrt{2}}{4}\, \big(A_{\mu p_1p_2} - B_\mu{}^q A_{qp_1p_2}\big) A_{p_3p_4p_5} \right) \notag \\[4mm]
\cB_{\mu\, m} &= -18\, {\eta}^{n_1\dots n_7} 
\Bigg( A_{\mu n_1 \dots n_7, m} + (3 \tilde{c}-1)
\left( A_{\mu  n_1 \dots n_5} - B_{\mu}{}^{p} A_{p n_1 \dots n_5}
\right) A_{n_6 n_7 m} \notag \\[2mm]
& \quad \, + \tilde{c} A_{ n_1 \dots n_6} \left( A_{\mu n_7 m} -
B_{\mu}{}^{p} A_{pn_7m} \right) + \frac{\sqrt{2}}{12} \left( A_{\mu n_1 n_2} -
B_{\mu}{}^{p} A_{p n_1 n_2} \right) A_{n_3 n_4 n_5} A_{n_6 n_7 m} \Biggr),
\label{Bdef1}
\end{align}
where $\tilde{c}$ is an undetermined constant.
These are related to the generalised vielbein via the following supersymmetry transformation \cite{dWNsu8, dWN13, GGN13}
\begin{equation}
 \delta \cB_{\mu}{}^{\cM} = i \, \Omega^{\cM \cN} \cV_{\cN \, AB} \left(2 \sqrt{2} \overline{\varepsilon}^{A}\varphi_{\mu}^{B} + \overline{\varepsilon}_{C} \gamma_{\mu} \chi^{ABC} \right) \,+ \, \textup{h.c.}
\end{equation}
using the supersymmetry transformations of the fields given in \cite{dWNsu8, dWN13, GGN13}. In particular \cite{GGN13}
\begin{align}
 \delta A_{\mu m_1 \dots m_7, n} &= -\frac{1}{9!} \left( \overline{\varepsilon} \tilde{\Gamma}_{\mu m_1 \dots m_7} \Psi_{n} - 8 \overline{\varepsilon} \tilde{\Gamma}_{n}  \tilde{\Gamma}_{[\mu m_1 \dots m_6} \Psi_{m_7]} \right) + \frac{\sqrt{2} \, \tilde{c}}{5!} \overline{\varepsilon} \tilde{\Gamma}_{[\mu m_1 \dots m_4} \Psi_{m_5} A_{m_6 m_7] n}  \notag \\[2mm]
 & \quad + \frac{\sqrt{2}}{3} \overline{\varepsilon} \tilde{\Gamma}_{[\mu m_1} \Psi_{m_2} \left( A_{m_3 \dots m_7]n} + \frac{\sqrt{2}}{12} A_{m_3 \dots m_5} A_{m_6 m_7]n} \right)  \notag \\[1mm]
 & \quad -   \sqrt{2} \, \tilde{c} \, \overline{\varepsilon} \tilde{\Gamma}_{[\mu m_1} \Psi_{m_2} \left( A_{m_3 \dots m_7]n} + \frac{\sqrt{2}}{4} A_{m_3 \dots m_5} A_{m_6 m_7]n} \right),  \label{susyh}
\end{align}
where $\Psi_{m}$ is the component of the $D=11$ gravitino along the internal directions
(prior to any redefinition).

The transformation of the components of $\cB_{\mu}^\cM$ under internal diffeomorphisms is
\begin{align}
 \delta \cB_{\mu}{}^{m} &= \xi^{p} \partial_{p} \cB_{\mu}{}^{m} \,-\, \partial_{p} \xi^{m} \cB_{\mu}{}^{p}, \notag  \\[2mm]
 \delta \cB_{\mu mn} &= \xi^{p} \partial_{p} \cB_{\mu mn} \,-\, 2 \, \partial_{[m} \xi^{p} \cB_{\mu n] p}, \notag \\[2mm]
 \delta \cB_{\mu}{}^{mn} &= \xi^{p} \partial_{p} \cB_{\mu}{}^{mn} \,+\, 2\, \partial_{p} \xi^{[m} \cB_{\mu}{}^{n]p} + \partial_{p} \xi^{p} \cB_{\mu}{}^{mn}. 
\label{cotransb}
 \end{align}
We note that $\cB_{\mu}{}^{mn}$ transforms as a tensor density of weight 1 because of the tensor density $\eta$ in its definition, \eqref{Bdef1}.
The transformation of $\cB_{\mu}^\cM$ 
under internal 2-form and 5-form gauge transformations is
\begin{align}
& \delta \cB_\mu{}^{m} = 0, \hspace{10mm} \delta \cB_{\mu mn} = - 36 \sqrt{2}\, \partial_{[m} \xi_{np]} \cB_\mu{}^{p} \notag \\[2mm]
& \delta \cB_\mu{}^{mn} = - 3\sqrt{2} \, \Delta \epsilon^{mnp_1\cdots p_5}\,
           \partial_{p_1} \xi_{p_2p_3}  \cB_{\mu p_4 p_5}. \label{dB2form}
\end{align}
and
\begin{eqnarray}
\hspace{10mm} \delta \cB_\mu{}^{m} = \delta \cB_{\mu mn}= 0, \qquad
\delta \cB_\mu{}^{mn} = - 6\cdot 6! \sqrt{2}  \, \eta^{mnp_1\cdots p_5} 
\partial_{[q} \xi_{p_1 \cdots p_5]} \cB_\mu{}^{q}.\label{dB5form}
\end{eqnarray}
Since we do not know at this point how $A_{\mu m_1 \dots m_7, n}$ transforms under coordinate, 2-form and 
5-form gauge transformation we cannot, yet, determine the gauge transformation rule for 
the final component $\cB_\mu{}_{m}$. Let us nevertheless anticipate the results of
section \ref{sec:gendiff}, where we will find the transformation rule from the E$_{7(7)}$ 
structure of internal coordinate and gauge transformations:
\begin{gather} \label{cotransb4}
\delta \cB_{\mu m} = \xi^{p} \partial_{p} \cB_{\mu m} \,+ \, \partial_{m} \xi^{p} \cB_{\mu p} + \partial_{p} \xi^{p} \cB_{\mu}{}^{mn},  \\[2mm]
\label{bdmu}
 \delta \cB_{\mu}{}_{m} = -18 \sqrt{2} \partial_{[m} \xi_{pq]} \cB_{\mu}{}^{pq}, \qquad  \delta \cB_{\mu}{}_{m} = 3\cdot 6! \sqrt{2} \, \eta^{n_1\cdots n_7} \partial_{[m} \xi_{n_1\cdots n_5]} 
   \cB_{\mu}{}_{n_6 n_7},
\end{gather}
for coordinate, 2-form and 5-form gauge transformations, respectively. Going backwards from these expressions, we can deduce that $A_{\mu n_1 \dots n_7,m}$ transforms as a tensor under internal coordinate transformations and 
under 2-form and 5-form gauge transformations it transforms as:
\begin{align}
 \delta A_{\mu n_1 \dots n_7,m} &= -18\, \tilde{c}\, \partial_{[m} \xi_{n_1 n_2]} A_{\mu n_3 \dots n_7} + \sqrt{2}(9 \tilde{c}-2)  \partial_{n_1} \xi_{n_2 n_3} A_{\mu n_4 n_5} A_{ m n_6 n_7} \notag \\[3mm] \label{a42g}
& \hspace{70mm} - \frac{(9 \tilde{c}-2)}{\sqrt{2}}  \partial_{[m} \xi_{n_1 n_2]} A_{\mu n_3 n_4} A_{n_5 \dots n_7}, \\
\delta A_{\mu n_1 \dots n_7,m} &= -6! \, (3 \tilde{c}-1) \partial_{[m} \xi_{n_1 \dots n_5]} A_{\mu n_6 n_7}, \label{a45g}
\end{align}
respectively. Here $\tilde c$ is the undetermined constant that appeared already 
in \cite{GGN13}, and that is also not fixed by imposing E$_{7(7)}$ covariance.
As for the generalised vielbein, we will show that the formulae  \eqref{dB2form},
\eqref{dB5form} and \eqref{bdmu}, together with the action of internal diffeomorphisms,
can be compactly assembled into a single E$_{7(7)}$ covariant formula, \eqref{genLieB}.

\section{Generalised vielbein postulate} \label{sec:GVP}

The generalised vielbeine satisfy differential constraints along the four external and 
the seven internal directions, which are called generalised vielbeine postulates (GVPs) 
in analogy with the usual vielbein postulate in differential geometry. These constraints
are identities that can be directly verified from the explicit expressions given above,
just like the usual vielbein postulate is an identity when the affine connection and the 
spin connection are expressed in terms of the usual vielbein. 

The external GVPs, which are the GVPs along the $d=4$ directions are of the form~\footnote{Note that the sign in front of the $\cP$ structures in both the external and internal GVPs is opposite to what appears in the GVPs as written in Ref.~\cite{GGN13}. This is because of a differing definition of the generalised vielbein $\cV$---more specifically, an extra factor of $i$ in the definition of $\cV$.}
\begin{align} \label{fgvp1}
& \partial_{\mu} \cV^{m}{}_{AB} + \cQ_{\mu}^{C}{}_{[A} \cV^{m}{}_{B]C} + 2 {\cB_\mu}^n D_n \cV^m{}_{AB} - 2  D_n \cB_{\mu}{}^m \cV^n{}_{AB} -D_n \cB_{\mu}{}^{n} \cV^{m}{}_{AB} = \cP_{\mu\, ABCD} \cV^{m\, CD}, \\[4mm] \label{fgvp2}
&\partial_\mu \cV_{mn\, AB} + \cQ_{\mu}^{C}{}_{[A} \cV_{|mn|\, B]C} + 2 {\cB_\mu}^p D_p \cV_{mn\, AB} - 4  D_{[m}{\cB_{|\mu|}}^p \cV_{n]p\, AB} - D_p \cB_{\mu}{}^{p} \cV_{mn \, AB} \notag \\[3mm]
&\hspace{88mm}   + 6 D_{[m}\cB_{|\mu|\, np]} \cV^p{}_{AB} =  \cP_{\mu\, ABCD} \cV_{mn}{}^{CD}, \\[5mm] \label{fgvp3}
&\partial_{\mu} \cV^{mn}{}_{AB} + \cQ_{\mu}^{C}{}_{[A} \cV^{mn}{}_{B]C} + 2 {\cB_{\mu}}^p D_p \cV^{mn}{}_{AB}  + 6 D_{p}\cB_{\mu}{}^{[m} \cV^{np]}{}_{AB} - D_p {\cB_{\mu}}^p  \cV^{mn}{}_{AB} \notag \\[2mm]
&\hspace{36mm} +\frac{1}{2} \eta^{mnp_1 \ldots p_5} D_{p_1} \cB_{\mu \, p_2 p_3} \cV_{p_4 p_5\, AB} + 4 D_p \cB_{\mu}{}^{p[m} \cV^{n]}{}_{AB} = \cP_{\mu\, ABCD} \cV^{mn\, CD}, \\[15pt] \label{fgvp4}
& \partial_\mu \cV_{m\, AB} + \cQ_{\mu}^{C}{}_{[A} \cV_{m\, B]C} 
 + 2 {\cB_{\mu}}^p D_p \cV_{m\, AB}  + 2 D_{m}\cB_{\mu}{}^{p} \cV_{p\, AB} + D_p {\cB_{\mu}}^p  \cV_{m\, AB} \notag \\[4mm]
&\hspace{58.5mm}
 +3 D_{[m} \cB_{|\mu| pq]} \cV^{pq}{}_{AB} - 2 D_{p} \cB_{\mu}{}^{pq} \cV_{qm\, AB} = \cP_{\mu\, ABCD} \cV_{m}{}^{CD},
\end{align}
where $D_m$ is the covariant derivative with respect to seven-dimensional 
diffeomorphisms, e.g.~
\begin{equation}
D_m \cB_\mu{}^n \equiv \partial_m \cB_\mu{}^n + \Gamma_{mp}^n \cB_\mu{}^p
\end{equation}
with the internal affine connection $\Gamma_{mn}^p$. In GVPs, above, the combination of components of the vector field $\cB_{\mu}{}^{\cM}$ and the generalised vielbein in each term is exactly such that the discrepancy in the weights of the components of the generalised vielbein is compensated by the differing weights in the components of the vector field $\cB_{\mu}{}^{\cM}.$ Hence the weights of the terms in each GVP are consistent. 

Note that in previous work \cite{dWNsu8, GGN13}
these relations were given without the affine connection terms, but the relations above
are still equivalent to the original ones (see \cite{dWNcc}), as all terms containing the affine connections 
cancel in the above relations, as well as the ones given below.
The connection coefficients are of the form
\begin{align}
 \cQ_{\mu}^{A}{}_{B} &= - \textstyle{\frac{1}{2}} \Big[ {e^m}_a D_{m} B_{\mu}{}^{n} e_{n b} - ({e^p}_{a} \cD_{\mu} e_{p\, b}) \Big] \Gamma^{ab}_{AB} 
 - \textstyle{\frac{\sqrt{2}}{12}} {e_{\mu}}{}^{\alpha} \left( F_{\alpha abc} \Gamma^{abc}_{AB} - \eta_{\alpha \beta \gamma \delta} F^{\beta \gamma \delta a} \Gamma_{a AB} \right), \\[3mm]
 \cP_{\mu ABCD}& = \textstyle{\frac{3}{4}} \Big[ {e^m}_a D_{m} B_{\mu}{}^{n} e_{n b} - ({e^p}_{a} \cD_{\mu} e_{p\, b}) \Big] \Gamma^{a}_{[AB} \Gamma^{b}_{CD]}
 - \textstyle{\frac{\sqrt{2}}{8}} {e_{\mu}}{}^{\alpha} F_{abc \alpha} \Gamma^{a}_{[AB} \Gamma^{bc}_{CD]} \notag \\[2mm]
& \hspace{75mm} - \textstyle{\frac{\sqrt{2}}{48}} e_{\mu \, \alpha} \eta^{\alpha \beta \gamma \delta} F_{a \beta \gamma \delta}{\Gamma_{b}}_{[AB} \Gamma^{ab}_{CD]},
\end{align}
where
\begin{equation}
\cD_\mu \equiv \partial_\mu - B_\mu{}^m D_m
\equiv \partial_\mu + 2 \cB_\mu{}^m D_m \, .
\end{equation}
In the dimensionally reduced theory, the kinetic term for the scalar fields is 
$\propto \cP_\mu^{ABCD} \cP^\mu_{ABCD}$, while the `composite' SU(8) connection
$\cQ_{\mu}{}^A{}_B$ is required for the covariantisation of the fermionic couplings.

Similarly, the generalised vielbein satisfies a GVP along the internal directions.
The relevant relations were derived in \cite{GGN13} and read
\begin{align}
  & \partial_{p} \cV^{m}{}_{AB} +  \Gamma_{pn}^m \cV^{n}{}_{AB} +  \frac12\Gamma_{pn}^n \cV^{m}{}_{AB} +\cQ_{p}^{C}{}_{[A} \cV^{m}{}_{B]C} = \cP_{pABCD} \cV^{m\, CD}, \label{7gvp1} \\[3mm]
 &\partial_{p} \cV_{mn\, AB} + 2\Gamma_{p[m}^q \cV_{n]q\, AB} +   \frac12 \Gamma_{pq}^q \cV_{mn\, AB}
    -  6\sqrt{2} \,  \Xi_{p|mnq} \cV^{q}{}_{AB} +  \cQ_{p}^{C}{}_{[A} \cV_{mn\, B]C} \notag \\[3mm]
&\hspace{111mm} =  \cP_{pABCD} \cV_{mn}{}^{CD}, \label{7gvp2} \\[5mm]
& \partial_{p} \cV^{mn}{}_{AB} - 2\Gamma_{pq}^{[m} \cV^{n]q}{}_{AB} - \frac12 \Gamma_{pq}^q \cV^{mn}{}_{AB} 
  - 6\sqrt{2} \eta^{mnq_1\cdots q_5}\,  \Xi_{p|q_1 \dots q_6} \cV^{q_6}{}_{AB} \, \notag   \\[2mm] 
 & \hspace{31mm}   - \frac{1}{\sqrt{2}}\, \eta^{mnq_1\cdots q_5}\,  \Xi_{p|q_1 q_2q_3} \cV_{q_4 q_5\,  AB} 
  + \cQ_{p}^{C}{}_{[A} \cV^{mn}{}_{B]C} = \cP_{pABCD} \cV^{mn\, CD}, \label{7gvp3} \\[5mm]
& \partial_p \cV_{m\, AB} - \Gamma_{pm}^n \cV_{n\, AB} - \frac12 \Gamma_{pq}^q \cV_{m\, AB}
- \sqrt{2} \eta^{n_1\cdots n_7}\, \Xi_{p|n_1\cdots n_6} \cV_{n_7 m\, AB} -  3\sqrt{2}\, \Xi_{p|rsm} \cV^{rs}{}_{AB} \notag \\[2mm]
& \hspace{89mm}  + \cQ_{p}^{C}{}_{[A} \cV_{m\, B]C} =  \cP_{pABCD} \cV_{m}{}^{CD}, \label{7gvp4}
\end{align}
where the first few terms in each of the above equations correspond to the general covariant derivative, i.e.
\begin{eqnarray}
D_m \cV^n_{AB} &\equiv& \partial_m \cV^n_{AB} \,+\, \Gamma_{mp}^n \cV^p_{AB} 
        \,+\, \frac12 \Gamma_{mp}^p e^n_{AB},\\
D_p \cV_{mnAB} &\equiv& \partial_p \cV_{mnAB} \, + \, 2 \Gamma_{p[m}^q \cV_{n]qAB} 
        \, + \, \frac12 \Gamma_{mp}^p e^n_{AB},
\end{eqnarray}        
and so on. Note that the components of the generalised vielbein are densities with respect to internal
coordinate transformations, hence the extra terms involving $\Gamma_{mn}^n$.
Furthermore, the connection coefficients $\cQ_{m}^{A}{}_{B}$ and $\cP_{mABCD}$ are 
\begin{gather}
 \cQ_{m}^{A}{}_{B} = - \textstyle{\frac{1}{2}} 
 \omega_{m\,ab} \Gamma^{ab}_{AB} + \textstyle{\frac{\sqrt{2}}{14}} i f e_{m a} \Gamma^{a}_{AB} - \textstyle{\frac{\sqrt{2}}{48}} e_{m}{}^{a} F_{abcd} \Gamma^{bcd}_{AB}, \\[3mm]
 \cP_{mABCD} = 
   \textstyle{\frac{\sqrt{2}}{56}} i f {e_{m}}^{a} {\Gamma_{ab}}_{[AB} \Gamma^{b}_{CD]} + \textstyle{\frac{\sqrt{2}}{32}} e_{m}{}^{a} F_{abcd} \Gamma^{b}_{[AB} \Gamma^{cd}_{CD]},
\end{gather}
where
\begin{equation}
f= -\textstyle{\frac{1}{24}} i \eta^{\alpha \beta \gamma \delta} F_{\alpha \beta \gamma \delta} = - \frac{1}{7!}  \eta^{a_1 \dots a_7} F_{a_1 \dots a_7}.
\end{equation}
The above connection coefficients can also be written in a more suggestive form 
\begin{gather}
 \cQ_{m}^{A}{}_{B} = - \textstyle{\frac{1}{2}} 
 \omega_{m\,ab} \Gamma^{ab}_{AB} + \textstyle{\frac{\sqrt{2}}{14 \cdot 6!}} F_{m a_1 \dots a_6} \Gamma^{a_1 \dots a_6}_{AB} - \textstyle{\frac{\sqrt{2}}{48}} F_{mabc} \Gamma^{abc}_{AB}, \\[3mm]
 \cP_{mABCD} = 
   - \textstyle{\frac{\sqrt{2}}{56 \cdot 5!}} F_{m a_1 \dots a_6} \Gamma^{a_1}_{[AB} \Gamma^{a_2 \dots a_6}_{CD]} + \textstyle{\frac{\sqrt{2}}{32}} F_{mabc} \Gamma^{a}_{[AB} \Gamma^{bc}_{CD]},
\end{gather}
whence it is clear that they are {\em invariant} under 2-form and 5-form gauge 
transformations. The expressions for $\cQ_{m}^{A}{}_{B}$ and $\cP_{mABCD}$ given here differ from the expressions given before~\footnote{See equations (3.33) 
and (3.34) of Ref.~\cite{dWNsu8}.} because of the replacement
\begin{equation}
{e^p}_{a} \partial_{m} e_{p\, b}    \; \rightarrow \;
{e^p}_{a} D_{m} e_{p\, b} \equiv - \omega_{m\,ab},
\end{equation}
where $D_{m} e_{n\, a}\equiv \partial_m e_{n\,a} - \Gamma_{mn}^p e_{p\,a}$, so that
$\omega_{m\,ab}$ is just the usual spin connection (these modifications to the GVP
were already introduced in \cite{dWNcc}). By contrast there is now no 
contribution to $\cP_{mABCD}$ from the derivative of the siebenbein because the 
spin connection is antisymmetric in $[ab]$ and thus vanishes when contracted 
with $\Gamma^a_{[AB} \Gamma^b_{CD]}$. Nevertheless, the GVPs are fully equivalent to the ones given previously, with the only difference being that some of the terms have now been absorbed into the affine connection terms.
One advantage of this rearrangement is that both $\cQ_m$ and $\cP_m$ now transform
as proper vectors under internal diffeomorphisms, unlike the expressions originally
given in \cite{dWNsu8}.

The essential new feature in the internal GVPs \eqref{7gvp1}--\eqref{7gvp4} is the 
appearance of new affine connection coefficients associated with the form fields, to wit,
\begin{align} \label{Xi1}
 \Xi_{p|mnq} &\equiv\,  D_{p} A_{mnq} - \frac{1}{4!} F_{pmnq}, \\[1mm]
 \Xi_{p|m_1 \cdots m_6} &\equiv \, D_{p} A_{m_1 \cdots m_6} + \frac{\sqrt{2}}{48} F_{p[m_1 m_2 m_3} A_{m_4 m_5 m_6]} \notag \\[3pt]
 & \qquad - \frac{\sqrt{2}}{2} \left( D_{p} A_{[m_1 m_2 m_3} - \frac{1}{4!} F_{p[m_1 m_2 m_3} \right) A_{m_4 m_5 m_6]} - \frac{1}{7!} F_{p m_1 \dots m_6}. \label{Xi2}
 \end{align}
 Observe that the above expressions vanish upon full antisymmetrisation:
 \begin{equation}
 \Xi_{[p|mnq]} = 0 \; ,\;\qquad
 \Xi_{[p|m_1 \cdots m_6]}  = 0
 \end{equation}
so that the gauge invariant 4-form and 7-form field strengths are uniformly projected out. 

Under 2-form and 5-form gauge transformations, respectively, the connections transform as
\begin{eqnarray}
 \delta \Xi_{p|mnq} &=& 3! D_{p} D_{[m} \xi_{nq]}, \notag\\[2mm]
\delta \Xi_{p|m_1 \cdots m_6} &=& 
-3! \, \sqrt{2} \left( D_{p} A_{[m_1 m_2 m_3} - \frac{1}{4!} F_{p[m_1 m_2m_3} \right) \partial_{m_4} \xi_{m_5m_6]},  \label{2gaugeXi}  
\end{eqnarray}
and
\begin{eqnarray}
 \delta \Xi_{p|mnq} &=& 0, \hspace{22.25mm}  \delta \Xi_{p|m_1 \cdots m_6} = 6! \, D_{p} D_{[m_1} \xi_{m_2 \cdots m_6]} \label{5gaugeXi} \;.
\end{eqnarray}
As expected, these transformations contain {\em second} derivatives of the transformation parameters, in complete analogy with the transformation of the usual affine connection under ordinary diffeomorphisms.

\section{Generalised diffeomorphisms} \label{sec:gendiff}

The E$_{7(7)}$ generalised Lie derivative \cite{CSW} incorporates the usual seven-dimensional 
spatial diffeomorphisms as well as the gauge transformations of the 3- and 6-form fields.
Indeed, it is immediately obvious from the explicit expressions \eqref{gv11}--\eqref{gv14}
that the generalised vielbein is not invariant under such gauge transformations, and
that the transformation properties can be read off directly from the components, equations \eqref{dV2form} and \eqref{dV5form}. We will therefore combine all these transformations and 
internal diffeomorphisms into a generalised Lie derivative 
of the 56-bein $\cV$, such that
\begin{equation} \label{gentrans}
 \delta_{\Lambda} \cV_{\cM\, AB} \,=\,  \hat{\cL}_{\Lambda} \cV_{\cM\, AB},
\end{equation}
where \cite{CSW, BCKT}~\footnote{The first prefactor is introduced for convenience:
$\frac12 \Lambda^\cM\partial_\cM \equiv \Lambda^{m8} \partial_{m8} + \cdots 
\equiv \Lambda^m \partial_m + \cdots.$} 
\begin{equation} \label{genLie}
 \hat{\cL}_{\Lambda} X_{\cM} \,=\,  \frac{1}{2} \Lambda^{\cN} \partial_{\cN} X_{\cM} 
 \,+ \, 6 (t^\alpha)_{\cM}{}^{\cN} (t_\alpha)_{\cP}{}^{\cQ} \partial_{\cQ} \Lambda^{\cP} X_{\cN}
\end{equation}
as is usually done in generalised geometry \cite{BGGP, CSW, BCKT, CSW2}. 
For a generalised covector density of weight $w,$ this formula generalises to
\begin{equation} \label{genLiew}
 \hat{\cL}^{{\scriptscriptstyle{(w)}}}_{\Lambda} X_{\cM} \,=\, \frac{1}{2} \Lambda^{\cN} \partial_{\cN} X_{\cM} 
 \,+\,  6 (t^\alpha)_{\cM}{}^{\cN} (t_\alpha)_{\cP}{}^{\cQ} \partial_{\cQ} \Lambda^{\cP} X_{\cN} 
 \,+\, \frac{1}{2} w \, \partial_{\cN} \Lambda^{\cN} X^{\cM}.
\end{equation}
Thus the generalised vielbein $\cV$ has weight zero. These relations are very suggestive 
of 56 internal coordinates $y^\cM$, rather than only
the seven internal coordinates $y^m$ coming from $D=11$ supergravity. However, 
it should be understood that all relations are valid {\em only in conjunction with the section condition}~\footnote{The section condition first appeared in the context of O$(d,d)$ \cite{Siegel:1993th}, where it is equivalent to the level-matching condition in bosonic string theory.} \cite{CSW, BCKT}
\begin{equation}
t_\alpha^{\cM\cN}\, \partial_\cM \otimes \partial_\cN = 0, \qquad \Omega^{\cM\cN}\, \partial_\cM \otimes \partial_\cN = 0.
\label{seccon}
\end{equation}
This condition is crucial in order for the algebra of generalised gauge transformations to be properly defined. In fact, the closure of the algebra and the Jacobi identity only hold if the above condition is satisfied \cite{Hohm:2010pp, BGGP, CSW, BCKT}. This condition also allows one to introduce the extra structure associated with extra coordinates, and in an E$_{7(7)}$
covariant manner, without having to view eleven-dimensional supergravity as a 
\emph{bona fide} Kaluza-Klein reduction of yet another genuinely higher dimensional 
theory (which does not appear to exist). Therefore, the requirement that 
\begin{equation}
 \partial_{\cM} = \begin{cases}
                   \partial_{m}  \quad & \text{if $\cM = m8$}, \\
		    0 & \text{otherwise}
                  \end{cases}
\label{secconsol}
\end{equation}
is not a reduction ansatz, but simply a solution of constraint \eqref{seccon}. Another solution of the section condition leads to type IIB theory \cite{Hohm:2013pua, BMP, HO6, HO}. In this way by extending the coordinates one can unify these various descriptions in a single framework. In this work, we will always assume equation \eqref{secconsol}, reducing to a generalised geometric framework in the sense of Hitchin and Gualtieri \cite{Hitchin, Gualtieri}.    

In order to see the link with the explicit formulae at the end of section~2.1, we 
now decompose the gauge transformation parameter in terms of the GL(7) 
subgroup as follows:
\begin{equation}
 \Lambda^{\cM}=(\Lambda^{m}, \Lambda_{mn}, \Lambda^{pq}, \Lambda_{p}),
\end{equation}
Clearly, we can identify $\Lambda^m$ as the diffeomorphism parameter 
and $\Lambda_{mn}$ as the gauge parameter of three-form gauge transformations.  
Dualising $\Lambda^{pq}$ to a 5-form allows us to identify this as the gauge parameter for 6-form transformations.
The final component $\Lambda_{p} \equiv \Lambda_{p8} = - \Lambda_{8p}$ is less understood.  However, it is clearly related to gauge transformations associated with dual gravity.  Although, we do not have a good understanding 
(at least not beyond the linearised level)
of what these gauge transformations could involve, this does not cause us any problems.  This is because dual gravity degrees of freedom do not contribute to the 56-bein.  Equivalently, the 70 scalars in the four-dimensional theory have no contribution from the dualisation of gravitational degrees of freedom.  Therefore, we would expect that the generalised gauge transformation of the 56-bein with respect to $\Lambda_p$ transformations vanishes. This can be shown simply:
assume that 
\begin{equation*}
 \Lambda^{\cM}=(0,0,0, \Lambda_{p}).
\end{equation*}
Using equation \eqref{genLie}, the generalised Lie derivative of the 56-bein reduces to 
\begin{equation}
 \hat{\cL}_{\Lambda} \cV_{\cM\, AB} = 24 (t^\alpha)_{\cM}{}^{\cN} (t_\alpha)^{p8\, q8} \partial_{q} \Lambda_{p} \cV_{\cN\, AB}. \label{LieD}
\end{equation}
But, we know that
\begin{equation*}
 (t_\alpha)^{p8\, q8} = (t_\alpha)^{[p8\, q8]}=0.
\end{equation*}
Hence, the 56-bein does not transform with respect to $\Lambda_p$ gauge transformations. 

Similarly, it is straightforward to check that \eqref{gentrans} precisely reproduces the coordinate and gauge transformations, \eqref{Vdiff}, \eqref{dV2form} and \eqref{dV5form} with
\begin{equation} \label{Lambdaxi}
\Lambda^m = \xi^m, \qquad \;\; \Lambda_{mn} = 12\sqrt{2} \, \xi_{mn}, \;\; \qquad
\Lambda^{mn} =  6!\sqrt{2} \eta^{mnp_1\cdots p_5} \, \xi_{p_1\cdots p_5},
\end{equation}
where $\xi_{mn}$ and $\xi_{m_1 \ldots m_5}$ are the 2-form and 5-form gauge parameters, see equations \eqref{2formgauge} and \eqref{5formgauge}. As an example, consider the transformation of component $\cV_{mn\, AB}$:
\begin{align}
 \delta_{\Lambda}\cV_{mn \, AB} &= \hat{\cL}_{\Lambda} \cV_{mn \, AB} = \frac{1}{2} \Lambda^{\cN} \partial_{\cN} \cV_{mn\, AB} + 6 (t^\alpha)_{mn}{}^{\cN} (t_\alpha)_{\cP}{}^{\cQ} \partial_{\cQ} \Lambda^{\cP} \cV_{\cN\, AB} \notag \\[2mm]
&= \Lambda^{p} \partial_{p} \cV_{mn\, AB} + 12 (t^\alpha)_{mn}{}^{\tR\tS} (t_\alpha)_{\tP\tQ}{}^{q8} \partial_{q} \Lambda^{\tP\tQ} \cV_{\tR\tS\, AB} + 12 (t^\alpha)_{mn\, \tR\tS} (t_\alpha)^{\tP\tQ\, q8} \partial_{q} \Lambda_{\tP\tQ} \cV^{\tR\tS}{}_{AB} \notag \\[3mm]
&= \Lambda^{p} \partial_{p} \cV_{mn\, AB} - 4 \left(\delta^{\tR\tS}_{\tQ[n} \delta^{\, q8}_{m]\tP}-1/8\, \delta^{\tR\tS}_{mn} \delta^{q8}_{\tP\tQ}\right)\partial_{q} \Lambda^{\tP\tQ} \cV_{\tR\tS\, AB} + 6\delta_{mn\tR\tS}^{\tP\tQ\,  q8} \partial_{q} \Lambda_{\tP\tQ} \cV^{\tR\tS}{}_{AB} \notag \\[1mm]
&= \left(\xi^{p} \partial_{p} \cV_{mn\, AB} + 2 \partial_{[m} \xi^{q} \cV_{q]n\, AB} - \frac{1}{2} \partial_{p} \xi^{p} \cV_{mn\, AB}\right)  + 36 \sqrt{2} \partial_{[p} \xi_{mn]} \cV^{p}{}_{AB},
\end{align}
where we have made use of the E$_{7(7)}$ representation given in appendix \ref{app:e7}.  As can be verified by a direct computation using the definition of $\cV_{mn\, AB}$ given in \eqref{gv12}, this corresponds to the transformation of $\cV_{mn\, AB}$ under coordinate and 3-form gauge transformations, \eqref{dV2form}. Verifying the precise agreement between \eqref{gentrans} and the 
formulae derived in section~2.1 for the remaining components is equally straightforward.

We require that the generalised Lie derivative $\hat{\cL}$ satisfies the product rule and that a generalised scalar transforms as
\begin{equation}
 \hat{\cL}_{\Lambda} S = \frac{1}{2} \Lambda^{\cN} \partial_{\cN} S,
\end{equation}
from which the Lie derivative of any generalised tensor can be found. In particular,
\begin{equation} \label{genLie2}
 \hat{\cL}_{\Lambda} X^{\cM} = \frac{1}{2} \Lambda^{\cN} \partial_{\cN} X^{\cM} - 6 (t^\alpha)_{\cN}{}^{\cM} (t_\alpha)_{\cP}{}^{\cQ} \partial_{\cQ} \Lambda^{\cP} X^{\cN}.
\end{equation}
This encodes the gauge transformations of the components of vector fields $\cB_{\mu}^{\cM},$ equations \eqref{dB2form} and \eqref{dB5form}. However, in order to obtain the correct coordinate transformations, \eqref{cotransb} and \eqref{cotransb4}, one must identify $\cB_{\mu}^{\cM}$ as a generalised tensor density of weight 1/2, i.e.\  
\begin{equation} \label{genLieB}
 \hat{\cL}_{\Lambda} \cB_{\mu}^{\cM} = \frac{1}{2} \Lambda^{\cN} \partial_{\cN} \cB_{\mu}^{\cM} - 6 (t^\alpha)_{\cN}{}^{\cM} (t_\alpha)_{\cP}{}^{\cQ} \partial_{\cQ} \Lambda^{\cP} \cB_{\mu}^{\cN} + \frac{1}{4} \partial_{\cN} \Lambda^{\cN} \cB_{\mu}^{\cM}.
\end{equation}

Furthermore, we can use equation \eqref{genLieB} to find the transformation of $\cB_{\mu}{}_{m}$ under 2-form and 5-form gauge transformations, equation \eqref{bdmu}.  Thus, we can deduce that the transformation of $A_{\mu n_1 \dots n_7,m}$ under 2-form and 5-form gauge transformations is given by equations \eqref{a42g} and \eqref{a45g}, respectively.

\section{Generalised vielbein postulates and generalised geometry} \label{sec:gvps}

The external GVPs, equations \eqref{fgvp1}--\eqref{fgvp4}, can be identified as the 
components of a single equation satisfied by $\cV$~\footnote{Recall that throughout 
this paper, we assume a solution of the E$_{7(7)}$ section condition of the form
\begin{equation*}
 \partial_{\cM} \neq 0 \quad\text{for $\cM = m$} \; , \qquad \text{otherwise 0}. 
\end{equation*}}
\begin{equation} \label{11gvp}
 \partial_\mu \cV_{\cM \, AB} + 2 \hat{\cL}_{\cB_{\mu}} \cV_{\cM\, AB} + \cQ_{\mu}^{C}{}_{[A} \cV_{\cM\, B]C} = \cP_{\mu\, ABCD} \cV_{\cM}{}^{CD}, 
\end{equation}
where $\hat{\cL}$ is the E$_{7(7)}$ generalised Lie derivative defined in 
equation \eqref{genLie}, or more specifically,
\begin{equation}
 \hat{\cL}_{\cB_{\mu}} \cV_{\cM\, AB}  \, = \, \frac12 \cB_\mu{}^\cN \partial_\cN \cV_{\cM AB} \, + \,
6 (t^\alpha)_{\cM}{}^{\cN} (t_\alpha)_{\cP}{}^{\cQ} \partial_{\cQ} \cB_\mu{}^{\cP} 
\cV_{\cN AB} \; .
\end{equation}
We note that the combination
\begin{equation*}
 \partial_\mu +2 \hat{\cL}_{\cB_{\mu}}  
\end{equation*}
already appears in reference \cite{HO} (see (2.27) of Ref.~\cite{HO}), where it is introduced as the covariant derivative with respect to $x$-dependent generalised diffeomorphisms. It is 
now straightforward to check that \eqref{11gvp} indeed coincides component by component
with equations \eqref{fgvp1}--\eqref{fgvp4}.

In a four-dimensional maximal gauge theory the scalars satisfy a Cartan equation of the form \cite{dWSTmax4}
\begin{equation} \label{4gvp}
 \partial_\mu \cV_{\cM \, ij} - g \cB_{\mu}{}^{\cP} X_{\cP \cM}{}^{\cN} \cV_{\cN ij}
  + \cQ_{\mu}^{k}{}_{[i} \cV_{\cM\, j]k}   \cV_{\cN\, ij} = \cP_{\mu\, ijkl} \cV_{\cM}{}^{kl}, 
\end{equation}
where $X_{\cM}$ generate the gauge algebra.
Comparing the eleven-dimensional equation \eqref{11gvp} with the four-dimensional equation \eqref{4gvp} to which it reduces under reduction, we find that from an eleven-dimensional point of view, the generators of the gauge algebra can schematically be viewed as a {\em differential operator} of the form~\footnote{We would like to thank Henning Samtleben for discussions on this.}
\begin{equation}
X_{\cP \cM}{}^{\cN}  \, = \, 
- \delta_{\cM}^{\cN} \overset{{}_{\shortrightarrow}}{\partial}_{\cP} - 12 (t^\alpha)_{\cM}{}^{\cN} (t_\alpha)_{\cP}{}^{\cQ} \overset{{}_{\shortleftarrow}}{\partial}_{\cQ}.
\end{equation}

The internal GVPs, \eqref{7gvp1}--\eqref{7gvp4}, likewise can be viewed as defining 
generalised connections: they can be written compactly as
\begin{equation} \label{intgvp}
 \partial_m \cV_{\cM\, AB} \, - \, {\bf{\Gamma}}_m{}_{\cM}{}^{\cN} \cV_{\cN\, AB} 
 \,+\,  \cQ_{m}^{C}{}_{[A} \cV_{\cM\, B]C} \,=\,  \cP_{m\, ABCD} \cV_{\cM}{}^{CD} \, .
\end{equation}
In this case, the generalised affine connection $\bGa_{\cM\cN}{}^\cP$ is non-zero only for the components with
$\cM = m$, which is the component appearing in equation \eqref{intgvp}.
With this restriction, it can be decomposed as
\begin{equation} \label{gammdef}
{\bf{\Gamma}}_m{}_{\cM}{}^{\cN} = {\bf{\Gamma}}_m{}^\alpha (t_\alpha)_{\cM}{}^{\cN}, 
\end{equation} 
It thus takes values in the Lie algebra of E$_{7(7)}$, in analogy with the usual affine connection 
that takes values in the Lie algebra of GL($n$); hence we can write
\begin{equation}
\qquad {\bf{\Gamma}}_m{}^\alpha\,\equiv \,\Big\{({\bf{\Gamma}}_m)_\tM{}^\tN\,,
\, ({\bf{\Gamma}}_m)^{\tM\tN\tP\tQ}\Big\} \, .
\end{equation}
In particular, we also define
\begin{equation}
({\bf{\Gamma}}_m)_{\tM\tN\tP\tQ} \equiv \frac1{24} \, \epsilon_{\tM\tN\tP\tQ\tR\tS\tT\tU}
({\bf{\Gamma}}_m)^{\tR\tS\tT\tU}\, .
\end{equation}

The components of the generalised affine connection $\bGa_{m \, \cN}{}^\cP$ 
can be read off by direct comparison with equations \eqref{7gvp1}--\eqref{7gvp4};
the non-zero components are
\begin{gather}
 (\bGa_{m})_{p8}{}^{q8} = - (\bGa_{m})^{q8}{}_{p8} = \textstyle{\frac{1}{2}} \Gamma_{mp}^{q} + \frac{1}{4} \Gamma_{mn}^{n} \delta_{p}^{q}, \qquad
 (\bGa_{m})_{pq}{}^{rs} = - (\bGa_{m})^{rs}{}_{pq} = 2 \Gamma_{m[p}^{[r}\delta_{q]}^{s]} -\textstyle{\frac{1}{2}} \Gamma_{mn}^{n} \delta_{pq}^{rs}, \notag \\[3mm]
 (\bGa_{m})_{p8}{}^{rs} = - (\bGa_{m})^{rs}{}_{p8} = 3 \sqrt{2} \, \eta^{rs t_1 \cdots t_5} \, \Xi_{m|p t_1 \cdots t_5}, \notag \\[3mm]
 (\bGa_{m})_{pq\, r8} = (\bGa_{m})_{r8 \, pq} = 3 \sqrt{2}\, \Xi_{m|pqr}, \qquad
 (\bGa_{m})^{pq\, rs} = \textstyle{\frac{1}{\sqrt{2}}}\, \eta^{pqrs t_1 t_2 t_3}\, \Xi_{m|t_1 t_2 t_3}, \label{bfg}
\end{gather}
where $\Gamma_{mn}^p$ is the usual affine connection for the seven internal directions.  Equivalently, the non-vanishing components of ${\bf{\Gamma}}_{\cM}{}^\alpha$ are
\begin{gather}
 ({\bf{\Gamma}}_m)_n{}^p \equiv - \Gamma_{mn}^p + 
 \textstyle\frac14 \delta_n^p \Gamma_{mq}^q, \qquad ({\bf{\Gamma}}_m)_8{}^8 = 
 - \textstyle \frac34\, \Gamma_{mn}^n, \notag \\[1mm]
 ({\bf{\Gamma}}_{m})_{8}{}^{n} = \sqrt{2} \eta^{n p_1 \cdots p_6}\, \Xi_{m|p_1 \cdots p_6}, \qquad
 ({\bf{\Gamma}}_{m})^{n_1 \cdots n_4} =  
 \textstyle\frac{1}{\sqrt{2}} \eta^{n_1 \cdots n_4 p_1 p_2 p_3}\, \Xi_{m| p_1 p_2 p_3}.
\end{gather}
Note that as required by the SL(8) property of the indices, we have 
$({\bf{\Gamma}}_m)_\tM{}^\tM = 0$. 

From a generalised geometry viewpoint, there is in principle no reason why a generalised connection $\bGa_{\cM \cN}{}^{\cP}$ cannot be non-zero for other values of $\cM$ such that
\begin{equation} 
 \partial_{\cM} \cV_{\cN\, AB} \, - \, {\bf{\Gamma}}_{\cM\cN}{}^{\cP} \cV_{\cP\, AB} 
 \,+\,  \cQ_{\cM}^{C}{}_{[A} \cV_{\cN\, B]C} \,=\,  \cP_{\cM \, ABCD} \cV_{\cN}{}^{CD} \, .
\end{equation}
In our approach this choice is made for us by the equations that come from $D=11$ supergravity. However, we can ``excite'' the other components by redefining $\cQ_{\cM}$ and $\cP_{\cM}.$ From an eleven-dimensional perspective, this freedom is allowed because it leaves the fermion supersymmetry transformations unchanged \cite{NP}. In any case, $D=11$ supergravity leads us to conclude that any covariant derivative that acts on the generalised vielbein only has components along the usual seven-dimensional space.  We also note that E$_{7(7)}$ valuedness of 
the affine connection implies
\begin{equation}
\cD_\cM \Omega_{\cN\cP} = 0
\end{equation}

The transformation of the generalised affine connection under generalised diffeomorphisms is
\begin{equation} \label{gencontrans}
 \delta_{\Lambda} {\bf{\Gamma}}_{\cM}{}_{\cN}{}^{\cP} = (\hat{\cL}^{{\scriptscriptstyle{(-1/2)}}}_{\Lambda} {\bf{\Gamma}} ) _{\cM}{}_{\cN}{}^{\cP} + 6 (t^{\alpha})_{\cN}{}^{\cP} (t_{\alpha})_{\cQ}{}^{\cR} \partial_{\cM} \partial_{\cR} \Lambda^{\cQ}, 
\end{equation}
where  $\hat{\cL}^{{\scriptscriptstyle{(-1/2)}}}_{\Lambda} {\bf{\Gamma}}$ is the canonical generalised Lie derivative of ${\bf{\Gamma}}$ with weight $-1/2$ along $\Lambda,$ equation \eqref{genLiew}. The above transformation encodes the usual inhomogeneous transformation of the affine connection as well as the gauge transformations of $\Xi$, which include second
derivatives of the 2-form and 5-form gauge parameters, equations \eqref{2gaugeXi} and \eqref{5gaugeXi}. 

When viewed as an analogue of the vielbein postulate, the internal GVP, \eqref{intgvp}, furnishes an E$_{7(7)}$ and SU(8) covariant derivative along $y^{m}.$ The generalised affine connection transforms in exactly such a way, \eqref{gencontrans}, so that given a generalised vector density $X^{\cM}$ of weight $w$,
\begin{equation} \label{conderw}
 \cD_{\cM} X^{\cN} \equiv \partial_{\cM} X^{\cN} + {\bf{\Gamma}}_{\cM}{}_{\cP}{}^{\cN} X^{\cP} - \frac{2}{3} w \, {\bf{\Gamma}}_{\cP}{}_{\cM}{}^{\cP} X^{\cN}
\end{equation}
transforms as a generalised tensor density of weight $(w\!-\!1/2)$(note the order of indices
in the last term, which is ${\bf{\Gamma}}_{\cP}{}_{\cM}{}^{\cP}$, and not
${\bf{\Gamma}}_{\cM}{}_{\cP}{}^{\cP}$). Observe that the weight term in the covariant derivative of a generalised tensor density differs from the usual covariant derivative of a tensor density because of the way ${\bf{\Gamma}}_{\cP}{}_{\cM}{}^{\cP}$  transforms under generalised diffeomorphisms, equation \eqref{gencontrans}. The fact that the covariant derivative of a tensor density must itself be a tensor density with weight $1/2$ less than the weight of the original tensor must be true of any covariant derivative that is defined in E$_{7(7)}$ generalised geometry. To see why this is true consider the non-covariant terms in the transformation of $\partial_{\cM} X^{\cN}$, where $X^{\cM}$ is a generalised tensor:
\begin{equation}
 \frac{1}{2} \partial_{\cM} \Lambda^{\cQ} \partial_{\cQ} X^{\cN} \,- \, 6 (t^\alpha)_{\cM}{}^{\cQ} (t_\alpha)_{\cP}{}^{\cR} \partial_{\cR} \Lambda^{\cP} \partial_{\cQ} X^{\cN} \,- \, 6 (t^\alpha)_{\cQ}{}^{\cN} (t_\alpha)_{\cP}{}^{\cR} \partial_{\cM} \partial_{\cR} \Lambda^{\cP} X^{\cQ}.
\end{equation}
The third term in the expression above must be cancelled by an inhomogeneous term in the transformation of the connection. However, using \cite{HO}
\begin{equation} \label{e7id}
 (t^\alpha)_{\cM}{}^{\cN} (t_\alpha)_{\cP}{}^{\cQ} = \frac{1}{12} \delta^{\cQ}_{\cM} \delta_{\cP}^{\cN} + \frac{1}{24} \delta^{\cN}_{\cM} \delta^{\cQ}_{\cP} +  (t^\alpha)_{\cM \cP} \,  (t_\alpha)^{\cN \cQ} - \frac{1}{24} \Omega_{\cM \cP} \,  \Omega^{\cN \cQ}
\end{equation}
and the section condition \eqref{seccon}, the remaining terms give
\begin{equation}
- \frac{1}{4} \partial_{\cQ} \Lambda^{\cQ} \partial_{\cM} X^{\cN},
\end{equation}
hence the covariant derivative of $X^{\cM},$ $\cD_{\cM} X^{\cN}$, must have weight 
$-1/2$ less than $X^\cM$ itself.~\footnote{The argument is essentially the same if $X^{\cM}$ is a generalised tensor density.}

\section{Generalised E$_{7(7)}$ curvature}

The generalised covariant derivative defined above can be used to define a 
generalised curvature (generalised Riemann tensor) $\bR$, given by
\begin{equation} \label{comcovd}
 [\cD_{\cM}, \cD_{\cN}] X_{\cP} = \bR_{\cM \cN \cP}{}^{\cQ} X_{\cQ}.
\end{equation}
Note that because of the transformation property of the covariant derivative, the second covariant derivative acts on a generalised tensor density of weight $-1/2$ (assuming that $X^{\cP}$ is a generalised tensor, and thus of weight zero). Hence the generalised Riemann tensor is in fact a tensor density of weight $-1$.  Furthermore, using equation \eqref{conderw} 
\begin{equation}
3 \, \bGa_{[\cM \cN]}{}^{\cQ}\, \partial_{\cQ} X_{\cP} = \bGa_{\cQ [\cM}{}^{\, \cQ} \, \partial_{\cN]} X_{\cP}
\end{equation} 
(which follows directly from the explicit expressions for the components in \eqref{bfg}),
the fact that the covariant derivative modifies the weight of the generalised tensor is crucial
in cancelling 
\begin{equation}
 \bGa_{[\cM \cN]}{}^{\cQ} \partial_{\cQ} X_{\cP}
\end{equation}
from the commutator of the covariant derivatives in equation \eqref{comcovd}. It is important 
that the term above is cancelled because from equation \eqref{gencontrans} the 
antisymmetrisation (in $\cM$ and $\cN$) of the generalised connection is {\em not} covariant -- this is unlike ordinary differential geometry where the antisymmetrisation of an affine connection can be identified as a covariant torsion.   In fact, a generalised torsion $\bT_{\cM \cN}{}^{\cP}$, as defined by
\begin{equation}
 [\cD_{\cM}, \cD_{\cN}] S = \bT_{\cM \cN}{}^{\cP} \partial_{\cP} S
\end{equation}
for some scalar $S$, can simply be shown to vanish in our scheme. Hence, the absence of a torsion 
term on the right hand side of equation \eqref{comcovd}. Let us emphasize once again that 
it is the explicit knowledge of the affine connection coefficients in \eqref{bfg} that enables 
us to overcome and resolve this well known difficulty encountered in previous work, see e.g. \cite{cek}.

Computing the left hand side of equation \eqref{comcovd} using the definition of the covariant derivative \eqref{conderw}, gives the form of the Riemann tensor, which turns out to be analogous to the expression for the conventional Riemann tensor in terms of an affine torsion-free connection
\begin{equation} \label{genR}
 \bR_{\cM \cN  \cP}{}^{\cQ} = -2 \partial_{[\cM} {\bf{\Gamma}}_{\cN]}{}_{\cP}{}^{\cQ} + 2 {\bf{\Gamma}}_{[\cM|}{}_{\cP}{}^{\cR} {\bf{\Gamma}}_{|\cN]}{}_{\cR}{}^{\cQ}. 
\end{equation}
Since the generalised Riemann tensor is defined using covariant derivatives, it is by definition an object that transforms covariantly under generalised diffeomorphisms, up to weight terms.  However, in appendix \ref{app:riem}, we explicitly verify that it transforms covariantly under generalised diffeomorphisms as a generalised tensor density of weight $-1$.

The non-zero components of the generalised Riemann curvature $\bR_{\cM \cN \cP}{}^{\cQ}$ 
can be directly computed from equations \eqref{bfg}; they are
\begin{gather}
 \bR_{m8\, n8\, p8}{}^{q8} = - \bR_{m8\, n8}{}^{q8}{}_{p8} = \frac{1}{2} R_{mnp}{}^{q}, \notag \\[2mm]
 \bR_{m8\, n8\, pq}{}^{rs} = - \bR_{m8\, n8}{}^{rs}{}_{pq} = 2 R_{mn[p}{}^{[r}\delta_{q]}^{s]}, \notag \\[3mm]
 \bR_{m8\, n8\, p8}{}^{rs} = - \bR_{m8\, n8}{}^{rs}{}_{p8} = 2 \sqrt{2}\, \delta^{[r}_{p} \eta^{s]t_1 \cdots t_6} D_{[m} \Xi_{n]|t_1 \cdots t_6} \notag \\[3mm]
 \hspace{80mm}+ 6 \eta^{rst_1 \cdots t_5} \Xi_{[m||t_1 t_2 p} \Xi_{|n]|t_4 t_5 t_6}, \notag \\[3mm]
 \bR_{m8\, n8\, pq\, r8} = \bR_{m8\, n8\, r8\, pq} = -6\sqrt{2} \, D_{[m}\Xi_{n]|pqr}, \notag \\[3mm]
 \bR_{m8\, n8}{}^{pq \, rs} = - \sqrt{2} \, \eta^{pqrst_1t_2 t_3} D_{[m}\Xi_{n]|t_1 t_2 t_3}. \label{Riemanncomp}
\end{gather}
Equivalently, decomposing $\bR$ as
\begin{equation}
 \bR_{\cM \cN \cP}{}^{\cQ} = \bR_{\cM \cN}{}^{\alpha} (t_{\alpha})_{\cP}{}^{\cQ}\;\, ,
\qquad \bR_{\cM\cN}{}^\alpha\,\equiv \,\Big\{(\bR_{\cM\cN})_\tM{}^\tN\,,
\, (\bR_{\cM\cN})^{\tM\tN\tP\tQ}\Big\} \, ,
\end{equation}
the non-zero components of $\bR_{\cM \cN}{}^{\alpha}$ are
\begin{gather}
 (\bR_{m8 \, n8})_{p}{}^{q} = - R_{mnp}{}^{q}, \notag \\[3mm]
(\bR_{m8 \, n8})_{8}{}^{p} =- 2 \eta^{pt_1 \cdots t_6} \left( \sqrt{2} D_{[m} \Xi_{n]|t_1 \cdots t_6} - \Xi_{m|t_1t_2t_3} \Xi_{n|t_4t_5t_6} \right), \notag \\[3mm]
 (\bR_{m8 \, n8})^{p_1 \cdots p_4} = -\sqrt{2} \eta^{p_1 \cdots p_4 rst} D_{[m} \Xi_{n]|rst}.
\end{gather}
Note that the above components of the generalised Riemann tensor are \emph{not} 
invariant under 2- and 5-form gauge transformations.  Indeed, this is to be expected since the definition of the generalised Riemann tensor as a generalised tensor ensures its \emph{covariance}, rather than invariance, under generalised gauge transformations.  While this may be antithetical to our usual notions of gauge transformations and how physical fields must accordingly transform, in a generalised geometric setting the appearance of gauge non-invariant terms should not come as a surprise, and one ought to view gauge transformations as being similar to coordinate transformations for which the notion of covariance, as well as invariance, exists.  In fact, we have already encountered this novelty before in the definitions of the generalised vielbein $\cV_{\cM\, AB}$ and vectors $\cB_{\mu}{}^{\cM}$.  However, ``gauge covariance'' limits the dependence of gauge non-invariant terms in a generalised tensor to bare 
3-form and 6-form potentials, or their gauge invariant field strengths.
Therefore, the gauge potentials can only enter gauge non-invariant terms without any 
derivatives, as in the E$_{7(7)}$ 56-bein, and the fact that they do (and thus all non-covariant 
terms cancel in the expressions below) constitutes a non-trivial consistency check
of our scheme. This claim can be explicitly verified for the generalised Riemann tensor by expressing the above components directly in terms of the 3-form and 6-form gauge 
potentials and their associated field strengths, using \eqref{Xi1} and \eqref{Xi2},
\begin{align}
& (\bR_{m8 \, n8})_{8}{}^{p} = - \, \eta^{pq_1 \cdots q_6} \left( 6 \sqrt{2} R_{mnq_1}{}^{r} A_{r q_2 \cdots q_6} -\frac{2\sqrt{2}}{7!} D_{[m}F_{n]q_1 \cdots q_6} + \frac{1}{6} D_{[m}F_{n]q_1q_2 q_3} A_{q_4 q_5 q_6} \right. \notag \\[3mm]
&\hspace{70mm}  - 3 R_{mnq_1}{}^{r}A_{rq_2 q_3} A_{q_4 q_5 q_6} - \frac{2}{(4!)^2} F_{m q_1 q_2 q_3} F_{n q_4 q_5 q_6} \Biggr), \notag \\[3mm] \label{Rcom2}
& (\bR_{m8 \, n8})^{p_1 \cdots p_4} = -\frac{3\sqrt{2}}{2} \eta^{p_1 \cdots p_4 rst} \left(R_{mnr}{}^{u} A_{ust} - \frac{1}{36} D_{[m}F_{n]rst} \right).
\end{align}

It can now be explicitly verified that the transformation of components of the generalised 
Riemann tensor under coordinate and gauge transformations precisely matches the 
transformation given by generalised diffeomorphisms, as expected. For example, consider the transformation of the following component of $\bR_{\cM \cN \cP}{}^{\cQ}$:
\begin{equation*}
 \bR_{m8\, n8}{}^{pq\, rs},
\end{equation*}
given in \eqref{Riemanncomp}. We find that its transformation as derived from 
the generalised Lie derivative is
\begin{equation}
\delta \bR_{m8\, n8}{}^{pq\, rs}=
 \cL^{\scriptscriptstyle (+1)}_{\Lambda}(\bR_{m8\, n8}{}^{pq\, rs}) - \frac{3}{4} \eta^{pqrs t u_1 u_2} R_{mnt}{}^{u_3} \partial_{[u_1}\Lambda_{u_2 u_3]}.
\end{equation}
Hence, while the generalised curvature tensor is a generalised density of weight $-1,$ the component above transforms as a tensor density of weight +1 under usual coordinate transformation and it also transforms non-trivially under a 2-form gauge transformation.  Noting that this component is equal to the expression given in equation \eqref{Rcom2} and using equation \eqref{Lambdaxi}, which gives the relation between $\Lambda_{mn}$ and the 2-form gauge transformation parameter $\xi_{mn}$, we find a precise match.

Now, consider the contraction of the generalised Riemann curvature:
\begin{equation}
 \bR_{\cM \cN} = \bR_{\cM \cP \cN}{}^{\cP}.
\end{equation}
It is simple to see that the only non-zero component of $\bR_{\cM \cN}$ is
\begin{equation}
 \bR_{m8 \, n8} = R_{mn}.
\end{equation}
Hence the only gauge-invariant objects that can be formed from the generalised Riemann tensor are the Ricci tensor and the Ricci scalar. Therefore, at the 2-derivative level the internal Ricci scalar can be obtained from the generalised Riemann tensor and the flux terms correspond to the trace of the square of $\cP_{m}.$ Together these would correspond to the potential. In this sense the way the potential would be written here is different to the approach in \cite{BGPW} where the potential is written as a sigma-model in terms of the generalised metric
$$ 
M_{\cM \cN} \,= \, \cV_{\cM}{}^{AB} \cV_{\cN\, AB} \,+ \, \cV_{\cN}{}^{AB} \cV_{\cM\, AB}  
$$
and both the Ricci scalar and the flux terms arise from the same terms. We defer a discussion of how the generalised geometry determines the potential in terms of the E$_{7(7)}$ structures given here to a future work.

\section{Discussion}

The issue of defining generalised differential geometric structures, such as connections and curvatures, associated with exceptional duality groups and using them to construct the dynamics is clearly an important one and has also been considered in Refs.~\cite{CSW,  Park:2013gaj, Aldazabal:2013mya, cek}.

In this paper, we use the formalism developed in Ref.~\cite{GGN13} to derive E$_{7(7)}$ generalised geometric structures, including generalised connections and curvatures as 
well as making explicit the higher dimensional origin of the embedding tensor, for 
which the GVP plays a central role.  We derive the E$_{7(7)}$ connection 
which is used to construct the generalised curvature tensor, from the internal GVP and ultimately the $D=11$ theory. Importantly, and apart from the generalisation of the affine 
and spin connections, the internal GVP  is not just of the form $D\cV = 0$, but has 
an extra contribution from $\cP_{m}$ (this vanishes in the 
absence of the 4-form and 7-form field strengths, however, and then the GVP
reduces to the standard one). This
is a main difference with the ansatz made in Ref.~\cite{cek}. Another notable feature of the 
generalised covariant derivative defined here is that it changes the weight of the 
resulting generalised tensor, which is crucial, from our perspective, in allowing 
a generalised Riemann tensor to be defined. Here again, we differ from previous work
which encountered difficulties in defining a generalised Riemann tensor.

In general, the approach taken in Refs.~\cite{CSW,  Park:2013gaj, Aldazabal:2013mya, cek} is to try to generalise geometric structures to exceptional geometry using notions taken from usual differential geometry, such as metric compatibility of the connection, while incorporating the novelty of generalised geometry.  For example, in Ref.~\cite{CSW}, the index on the generalised connection that is associated with the derivative has components along extended tangent space directions, unlike the connection defined here---although this can be done in our case as well (see comments after equation \eqref{gencontrans}).  However, the difference in approach allows us to find a new contribution to what can be viewed as a vielbein compatibility of the connection, namely, $\cP_{m}$ as well as a generalised Riemann tensor that transforms covariantly under full generalised diffeomorphisms, as well as local SU(8) transformations.
 Such a tensor has been lacking in the generalised geometry literature associated to exceptional as well as O$(d,d)$ duality; see also \cite{Jeon:2010rw, Hohm:2010xe, Jeon:2011cn, Coimbra:2011nw, Hohm:2011si, BCMP}.

While, at the two-derivative level, the only scalar constructible from the generalised Riemann tensor reduces to the usual internal Ricci scalar, at a higher-derivative level, other scalars can be constructed that have explicit dependence on the gauge potentials.  These along with other scalars constructed from structures such as $\cP_{m}$ may help in providing an understanding of higher-derivative corrections from a generalised geometric perspective.  We will consider this possibility in the future. 

The analysis performed in this paper can also be straightforwardly applied to the $3+8$ 
split of $D=11$ supergravity, pertinent to the E$_{8(8)}$ duality group. Some preliminary 
results for this case have already been obtained in Refs.~\cite{KNS, GGN13}, and we hope to extend these partial results in a future work.  

\vspace{1cm}

\noindent{\bf Acknowledgements:} We are grateful to Henning Samtleben and Olaf Hohm 
for discussions and explanations of their recent work, especially \cite{HO}. We are also grateful to Martin Cederwall and Axel Kleinschmidt for discussions.  We would like to thank Henning Samtleben and ENS Lyon for hospitality.  H.G.~and M.G.~would like to thank the Max-Planck-Institut f\"{u}r Gravitationsphysik (AEI) and in particular H.N. for hospitality.  H.G.~and M.G.~are supported by King's College, Cambridge.  The work of H.N. is supported in part by the Gay-Lussac-Humboldt Prize.

\appendix

\section{E$_{7(7)}$ algebra and identities} \label{app:e7}

In this appendix, we list useful equations with regard to the SL(8) decomposition of the E$_{7(7)}$ algebra:
\begin{align}
 &(t^{\tM}{}_{\tN})^{\tP\tQ}{}_{\tR\tS} = 2 \left(\delta^{\tP\tQ}_{\tN[\tS} \delta^{\tM}_{\tR]} - \frac{1}{8} \delta^{\tM}_{\tN} \delta^{\tP\tQ}_{\tR\tS} \right), \qquad \qquad
(t^{\tM}{}_{\tN})_{\tR\tS}{}^{\tP\tQ} = - 2 \left(\delta^{\tP\tQ}_{\tN[\tS} \delta^{\tM}_{\tR]} - \frac{1}{8} \delta^{\tM}_{\tN} \delta^{\tP\tQ}_{\tR\tS} \right),\\
&(t_{\tP\tQ\tR\tS})^{\tT_1 \ldots \tT_4} = \delta^{\tT_1 \ldots \tT_4}_{\tP\tQ\tR\tS}, \hspace{34.5mm}  (t_{\tP\tQ\tR\tS})_{\tT_1 \ldots \tT_4} = \frac{1}{4!} \eta_{\tP\tQ\tR\tS \tT_1 \ldots \tT_4}, \\[2mm]
 &\kappa^{\tM}{}_{\tN},{}^{\tP}{}_{\tQ} = 12 \left(\delta^{\tM}_{\tQ} \delta^{\tP}_{\tN} - \frac{1}{8} \delta^{\tM}_{\tN} \delta^{\tP}_{\tQ} \right), \hspace{22mm} \kappa_{\tM\tN\tP\tQ,\tR\tS\tT\tU} = \frac{2}{4!} \eta_{\tM\tN\tP\tQ\tR\tS\tT\tU}, \\[2mm]
 &(\kappa^{-1})_{\tN}{}^{\tM},{}_{\tQ}{}^{\tP} = \frac{1}{12} \left(\delta^{\tM}_{\tQ} \delta^{\tP}_{\tN} - \frac{1}{8} \delta^{\tM}_{\tN} \delta^{\tP}_{\tQ} \right),\hspace{13mm} (\kappa^{-1})^{\tM\tN\tP\tQ,\tR\tS\tT\tU} = \frac{1}{2\cdot 4!} \eta^{\tM\tN\tP\tQ\tR\tS\tT\tU}.
\end{align}
where $\kappa$ is the Cartan Killing form on E$_{7(7)}$.

\section{Generalised covariance of the curvature tensor} \label{app:riem}

The generalised curvature tensor, defined in \eqref{genR}, is
\begin{equation}
 \bR_{\cM \cN  \cP}{}^{\cQ} = -2 \partial_{[\cM} {\bf{\Gamma}}_{\cN]}{}_{\cP}{}^{\cQ} + 2 {\bf{\Gamma}}_{[\cM|}{}_{\cP}{}^{\cR} {\bf{\Gamma}}_{|\cN]}{}_{\cR}{}^{\cQ}. 
\end{equation}
In this appendix we show that $\bR_{\cM \cN  \cP}{}^{\cQ}$ as defined above indeed transforms as a generalised tensor density of weight $-1$ under the transformation of the generalised connection, given in equation \eqref{gencontrans},
\begin{equation}
 \delta_{\Lambda} {\bf{\Gamma}}_{\cM}{}_{\cN}{}^{\cP} = (\hat{\cL}^{{\scriptscriptstyle{(-1/2)}}}_{\Lambda} {\bf{\Gamma}} ) _{\cM}{}_{\cN}{}^{\cP} + 6 (t^{\alpha})_{\cN}{}^{\cP} (t_{\alpha})_{\cQ}{}^{\cR} \partial_{\cM} \partial_{\cR} \Lambda^{\cQ}.
\end{equation}
Under generalised diffeomorphisms, the transformation of the generalised Riemann tensor is
\begin{align*}
\delta_{\Lambda} \bR_{\cM \cN  \cP}{}^{\cQ} = & -2 \partial_{[\cM} \delta_{\Lambda} {\bf{\Gamma}}_{\cN]}{}_{\cP}{}^{\cQ} + 2 \left( \delta_{\Lambda} {\bf{\Gamma}}_{[\cM|}{}_{\cP}{}^{\cR} \right) {\bf{\Gamma}}_{|\cN]}{}_{\cR}{}^{\cQ}  + 2 {\bf{\Gamma}}_{[\cM|}{}_{\cP}{}^{\cR} \left( \delta_{\Lambda} {\bf{\Gamma}}_{|\cN]}{}_{\cR}{}^{\cQ}  \right), \\[4mm]
= & -2 (\hat{\cL}^{{\scriptscriptstyle{(-1/2)}}}_{\Lambda} \partial \bGa )_{\cM \cN \cP}{}^{\cQ} + 12 (t^{\alpha})_{[\cM|}{}^{\cR} (t_{\alpha})_{\cS}{}^{\cT} \partial_{\cT} \Lambda^{\cS} \partial_{\cR} \bGa_{|\cN] \cP}^{\cQ} - \partial_{[\cM|} \Lambda^{\cR} \partial_{\cR} \Lambda_{|\cN] \cP}{}^{\cQ} \\[3mm]
& + \frac{1}{2} \partial_{[\cM|} \partial_{\cR} \Lambda^{\cR} \bGa_{|\cN] \cP}{}^{\cQ} + 12 (t^{\alpha})_{[\cM|}{}^{\cR} (t_{\alpha})_{\cS}{}^{\cT} \partial_{|\cN]} \partial_{\cT} \Lambda^{\cS} \bGa_{\cR \cP}^{\cQ} + 2  ( \hat{\cL}^{{\scriptscriptstyle{(-1)}}}_{\Lambda} \bGa \cdot \bGa )_{\cM \cN \cP}{}^{\cQ}, 
\end{align*}
where the generalised Lie derivatives 
$\hat{\cL}^{{\scriptscriptstyle{(-1/2)}}}_{\Lambda}$ on $\partial\bGa$ and 
$\hat{\cL}^{{\scriptscriptstyle{(-1)}}}_{\Lambda}$ on $\bGa$
are defined in analogy with \eqref{genLiew}. Further evaluation of this expression 
by means of  the E$_{7(7)}$ identity \eqref{e7id} and the section condition \eqref{seccon}
yields
\begin{align*}
\delta_{\Lambda} \bR_{\cM \cN  \cP}{}^{\cQ} = & -2 (\hat{\cL}^{{\scriptscriptstyle{(-1/2)}}}_{\Lambda} \partial \bGa )_{\cM \cN \cP}{}^{\cQ} + \frac{1}{2} \partial_{\cR} \Lambda^{\cR} \partial_{[\cM|} \bGa_{|\cN] \cP}^{\cQ} + 2 ( \hat{\cL}^{{\scriptscriptstyle{(-1)}}}_{\Lambda} \bGa \cdot \bGa )_{\cM \cN \cP}{}^{\cQ}, \\[4mm]
= & \, ( \hat{\cL}^{{\scriptscriptstyle{(-1)}}}_{\Lambda} \bR )_{\cM \cN \cP}{}^{\cQ},
\end{align*}
Therefore, the generalised Riemann tensor transforms as a generalised density of weight $-1$.

\newpage 

\bibliography{gendiff}
\bibliographystyle{utphys}
\end{document}